\documentclass[12pt]{article}
\usepackage{setspace}
\usepackage{epsfig}
\usepackage{graphicx}
\usepackage{amssymb}
\def\be{\begin{equation}}
\def\ee{\end{equation}}
\def\ba{\begin{array}}
\def\ea{\end{array}}
\def\beqn{\begin{eqnarray}}
\def\eeqn{\end{eqnarray}}

\def\bt{\begin{tabular}}
\def\et{\end{tabular}}
\def\bc{\begin{center}}
\def\ec{\end{center}}
\begin{document}

  \title{$~~~~~~~~$Implications of $\theta_{13}$ on Fritzsch-like lepton mass matrices}

 \author{Priyanka Fakay$^1$, Samandeep Sharma$^1$, Rohit Verma$^2$,  \\
 Gulsheen Ahuja$^{1*}$, Manmohan Gupta$^1$\\
\\
 {$^1$ \it Department of Physics, Centre of Advanced Study, P.U.,
 Chandigarh, India.} \\
 {$^2$ \it Rayat Institute of Engineering and Information Technology, Ropar, India.}\\
 {($^*$\it email: gulsheenahuja@yahoo.co.in)}}

 \maketitle
 \begin{abstract}

Implications of the lepton mixing angles, in particular of
$\theta_{13} $, have been investigated for minimal as well as
non-minimal Fritzsch-like textures for the case of Majorana
neutrinos. Both, in minimal texture (texture 6 zero lepton mass
matrices) and non-minimal textures (two cases of texture 5 zero
lepton mass matrices), inverted hierarchy and degenerate scenario
of neutrino masses have been ruled out. The implications of
$\theta_{13} $ have been investigated on the lightest neutrino
mass $m_{\nu_1} $ as well as the effective Majorana mass $\langle
m_{ee} \rangle$.
\end{abstract}

~~~Keywords: Lepton mass matrices, mixing angle $\theta_{13}$.

 ~~~PACS numbers: 12.15.Ff, 14.60.Pq
\\
\section{Introduction}
In the last few years, impressive advances have been made in
understanding the phenomenology of neutrino oscillations through
solar neutrino experiments\cite{davis}-\cite{sno}, atmospheric
neutrino experiments \cite{super}, reactor based experiments
\cite{kam},\cite{chooz} and accelerator based experiments
\cite{minos}-\cite{min}. The recent measurements of angle
$\theta_{13} $ \cite{min1}-\cite{reno} have undoubtedly improved
our understanding of the neutrino mixing matrix. Interestingly a
not so small value of $ \theta_{13}$ which is close to CHOOZ
\cite{chz} bound and to the Cabibbo angle, coupled with the fact
that $ \theta_{23}$ may not be maximal \cite{hall}, \cite{gouvea},
strongly indicates that there may not be any symmetry in the
leptonic sector \cite{alt}. These observations have also deepened
the mystery of flavor mixings as the pattern of quark and lepton
mixing angles now looks to be significantly different. Since the
mixing angles are related to the corresponding mass matrices,
therefore, formulating viable fermion mass matrices becomes all
the more complicated especially when quarks and leptons have to be
described in a unified framework.

In the absence of any compelling theory of flavor dynamics to
explain the fermion masses and mixings, the present day approaches
can be broadly categorized into `top-down' and `bottom-up'. The
top-down approach consists of formulating the fermion mass
matrices at Grand Unified Theories (GUT) scale based on certain
fundamental theoretical principles. However, in the absence of any
such viable approach, it is desirable to follow the bottom-up
approach. This approach consists of finding the phenomenological
fermion mass matrices which are in tune with the low energy data,
thereby providing hints for the formulation of mass matrices at
the GUT scale. As an example of this approach, texture specific
mass matrices, introduced implicitly by Weinberg \cite{wein} and
explicitly by Fritzsch \cite{fritzsch}, have played an important
role in understanding the pattern of quark and lepton mixing
phenomena \cite{mixings}.

Recently, in the light of observation of neutrino mixing angle $
\theta_{13}$, Fukugita  $\it{et~al.}$ \cite{fukugita} (henceforth
referred to as FSTY) have updated their previous analysis
\cite{fukuprevious} of texture 6 zero Fritzsch-like lepton mass
matrices for normal hierarchy of neutrino masses and have arrived
at some interesting conclusions. In particular, considering
Majorana neutrinos with normal hierarchy of neutrino masses as
well as the recent measurements of angle $\theta_{13} $ the
authors find the mass of the lightest neutrino  to be 0.7 - 2.1
meV. Further, the effective Majorana mass $\langle m_{ee}
\rangle$, appearing in the double beta decay, comes out to be
3.7-5.6 meV. Also, it has been shown that the  sum of the masses
of the three neutrinos is 61 $\pm $ 2 meV. Furthermore, FSTY
without getting into details, have also concluded that inverted
hierarchy and degenerate scenario of the neutrino masses are ruled
out.

Recently, in the case of quarks it has been shown \cite{neelu}
that not only Fritzsch-like but also non Fritzsch-like texture 6
zero mass matrices have been ruled out. Therefore, in case one has
to formulate quark and lepton mass matrices on the same footing,
as advocated by Smirnov \cite{smirnov}, perhaps there is a need to
go beyond texture 6 zero mass matrices even for the leptons.
Further, recent measurements of angle $\theta_{13}$ as well as
continuous refinements of the other two mixing angles and
corresponding mass square differences also provide the necessary
motivation to go beyond the minimal texture considered by FSTY .

The purpose of the present communication, on the one hand, is to
rule out inverted hierarchy and degenerate scenario of neutrino
masses corresponding to texture 6 zero lepton mass matrices. On
the other hand, keeping in mind that the texture 5 zero mass
matrices are not ruled out completely in the quark sector, we have
made an attempt to extend the analysis of FSTY to texture 5 zero
Fritzsch-like lepton mass matrices with an emphasis on the
implications of angle $\theta_{13}$. In particular, we have
carried out a detailed analysis pertaining to hierarchies of
neutrino masses as well as the implications of $\theta_{13}$ and
the other two mixing angles on the lightest neutrino mass for the
two possible cases of texture 5 zero lepton mass matrices.

The detailed plan of the paper is as follows. In Section
(\ref{form}), we detail the essentials of the formalism regarding
the texture specific mass matrices. Inputs used in the present
analysis have been given in Section (\ref{inputs}) and the results
and discussion have been presented in Section (\ref{cal}).
Finally, Section (\ref{summ}) presents a few comments and
summarizes our conclusions.

\section{Texture specific mass matrices and the PMNS matrix \label{form}}
To define the various texture specific cases considered here, we
begin with the modified Fritzsch-like mass matrices, e.g.,
 \be
 M_{l}=\left( \ba{ccc}
0 & A _{l} & 0      \\ A_{l}^{*} & D_{l} &  B_{l}     \\
 0 &     B_{l}^{*}  &  C_{l} \ea \right), \qquad
M_{\nu D}=\left( \ba{ccc} 0 &A _{\nu} & 0      \\ A_{\nu}^{*} &
D_{\nu} &  B_{\nu}     \\
 0 &     B_{\nu}^{*}  &  C_{\nu} \ea \right),
 \label{frzmm}
 \ee
$M_{l}$ and $M_{\nu D}$ respectively corresponding to Dirac-like
charged lepton and neutrino mass matrices. Both the matrices are
texture 2 zero type with $A_{l(\nu)}
=|A_{l(\nu)}|e^{i\alpha_{l(\nu)}}$
 and $B_{l(\nu)} = |B_{l(\nu)}|e^{i\beta_{l(\nu)}}$, in case these
 are symmetric then $A_{l(\nu)}^*$ and $B_{l(\nu)}^*$ should be
 replaced by $A_{l(\nu)}$ and $B_{l(\nu)}$, as well as
 $C_{l(\nu)}$ and $D_{l(\nu)}$ should respectively be defined as $C_{l(\nu)}
 =|C_{l(\nu)}|e^{i\gamma_{l(\nu)}}$ and $D_{l(\nu)}
 =|D_{l(\nu)}|e^{i\omega_{l(\nu)}}$.

The texture 6 zero matrices can be obtained from the above
mentioned matrices by taking both $D_l$ and $D_{\nu}$ to be zero,
which reduces the matrices $M_{l}$ and $M_{\nu D}$ each to texture
3 zero type. Texture 5 zero matrices can be obtained by taking
either $D_l=0$ and $D_{\nu}\neq 0$ or $D_{\nu}=0$ and $D_l \neq
0$, thereby, giving rise to two possible cases of texture 5 zero
matrices, referred to as texture 5 zero $D_l=0$ case pertaining to
$M_l$ texture 3 zero type and $M_{\nu D}$ texture 2 zero type and
texture 5 zero $D_{\nu}=0$ case pertaining to $M_l$ texture 2 zero
type and $M_{\nu D}$ texture 3 zero type.

To facilitate the understanding of inverted hierarchy case and its
relationship to the normal hierarchy case, we detail the
essentials of formalism connecting the mass matrix to the neutrino
mixing matrix. For detailed diagonalization procedure we refer the
readers to our earlier work \cite{group}, however, to make the
manuscript readable we outline the essentials. Texture specific
hermitian mass matrix
  $M_k$, where $k=l, \nu D$, can be expressed as \be M_k= P_k^{\dagger} M_k^r P_k \,,  \label{mk} \ee
where $M_k^r$ is a real symmetric matrix with real eigenvalues and
$P_k^{\dagger}$ and $P_k$ are diagonal phase matrices. The
 matrix $M_k^r$ is diagonalized by the orthogonal
transformation $O_k$ yielding \be M_k=Q_k O_k \xi_k M_k^{diag}
O_k^T P_k,\label{mkeq2} \ee wherein, to facilitate the
construction of diagonalizing transformations for different
hierarchies, we have introduced $\xi_k$ defined as $ {\rm diag}
(1,\,e^{i \pi},\,1)$ for the case of normal hierarchy and as $
{\rm diag} (1,\,e^{i \pi},\,e^{i \pi})$ for the case of inverted
hierarchy.

The charged leptons case is fairly straight forward, whereas in
the case of neutrinos, the diagonalizing transformation is
hierarchy specific as well as requires some fine tuning of the
phases of the right handed neutrino mass matrix $M_R$. Noting
$k=\nu D$ and using see-saw mechanism
 \be M_{\nu}=-M_{\nu D}^T\,(M_R)^{-1}\,M_{\nu D},
 \ee
we can write
\be
M_{\nu}=-P_{\nu D} O_{\nu D} M_{\nu D}^{diag} \xi_{\nu D} O_{\nu
D}^T Q_{\nu D}^T (M_R)^{-1} Q_{\nu D} O_{\nu D} \xi_{\nu D} M_{\nu
D}^{diag} O_{\nu D}^T P_{\nu D}. \ee Assuming fine tuning, the
phase matrices $Q_{\nu D}^T$ and $Q_{\nu D}$ along with $-M_R$ can
be taken as $m_R ~{\rm diag} (1,1,1)$ as well as using the
unitarity of $\xi_{\nu D}$ and orthogonality of $O_{\nu D}$, the
above equation can be expressed as
\be
M_{\nu}= P_{\nu D} O_{\nu D} \frac{(M_{\nu D}^{diag})^2}{(m_R)}
O_{\nu D}^T P_{\nu D}. \label{mnu} \ee

The lepton mixing matrix, in terms of the matrices used for
diagonalizing the mass matrices $M_l$ and $M_{\nu}$, can be easily
obtained and is expressed as
 \be
U =(Q_l O_l \xi_l)^{\dagger} (P_{\nu D} O_{\nu D}). \label{mix}
\ee Eliminating the phase matrix $\xi_l$ by redefinition of the
charged lepton phases, the above equation becomes
\be
 U = O_l^{\dagger} Q_l P_{\nu D} O_{\nu D} \,, \label{mixreal} \ee
where $Q_l P_{\nu D}$, without loss of generality, can be taken as
$(e^{i\phi_1},\,1,\,e^{i\phi_2})$, $\phi_1$ and $\phi_2$ being
related to the phases of mass matrices and can be treated as free
parameters.

To understand the relationship between diagonalizing
transformations for different hierarchies of neutrino masses and
for the charged lepton case, we present the diagonalizing
transformation $O_k$, whose first element can be written as
 \be  O_k(11) = {\sqrt
\frac{m_2 m_3 (m_3+m_2-D_{l(\nu)})}
     {(m_1+m_2+m_3-D_{l(\nu)})
(m_1-m_3)(m_1-m_2)} }~, \label{diaggen} \ee where $m_1$, $m_2$,
$m_3$ are eigenvalues of $M_k$. In the case of charged leptons,
because of the hierarchy $m_e \ll m_{\mu} \ll m_{\tau}$, the mass
eigenstates can be approximated respectively to the flavor
eigenstates, as has been considered by several authors
\cite{fuku,xingn}. In this approximation, $m_{l1} \simeq m_e$,
$m_{l2} \simeq m_{\mu}$ and $m_{l3} \simeq m_{\tau}$, one can
obtain the first element of the matrix $O_l$ from the above
element given in equation (\ref{diaggen}), by replacing $m_1$,
$m_2$, $m_3$ by $m_e$, $-m_{\mu}$, $m_{\tau}$, e.g.,
 \be  O_l(11) = {\sqrt
\frac{m_{\mu} m_{\tau} (m_{\tau}-m_{\mu}-D_l)}
     {(m_{e}-m_{\mu}+m_{\tau}-D_l)
(m_{\tau}-m_{e})(m_{e}+m_{\mu})} } ~. \ee

Equation (\ref{diaggen}) can also be used to obtain the first
element of diagonalizing transformation for Majorana neutrinos,
assuming normal hierarchy, defined as $m_{\nu_1}<m_{\nu_2}\ll
m_{\nu_3}$, and also valid for the degenerate case defined as
$m_{\nu_1} \lesssim m_{\nu_2} \sim m_{\nu_3}$, by replacing $m_1$,
$m_2$, $m_3$ by $\sqrt{m_{\nu 1} m_R}$, $-\sqrt{m_{\nu 2} m_R}$,
$\sqrt{m_{\nu 3} m_R}$, e.g., \be O_{\nu}(11) = {\sqrt
\frac{\sqrt{m_{\nu_2}}
    \sqrt{m_{\nu_3}}
( \sqrt{m_{\nu_3}}-\sqrt{ m_{\nu_2}}-D_{\nu})}
{(\sqrt{m_{\nu_1}}-\sqrt{m_{\nu_2}} + \sqrt{m_{\nu_3}}- D_{\nu})
(\sqrt{m_{\nu_3}}-\sqrt{m_{\nu_1}}) (\sqrt{m_{\nu_1}} +
\sqrt{m_{\nu_2}} )} } \label{omajnh}, \ee where $m_{\nu_1}$,
$m_{\nu_2}$ and $m_{\nu_3}$ are neutrino masses. The parameter
$D_{\nu}$ is to be divided by $\sqrt{m_R}$, however as $D_{\nu}$
is arbitrary therefore we retain it as it is.

In the same manner, one can obtain the elements of diagonalizing
transformation for the inverted hierarchy case, defined as
$m_{\nu_3} \ll m_{\nu_1} < m_{\nu_2}$, by replacing $m_1$, $m_2$,
$m_3$ in equation (\ref{diaggen}) with $\sqrt{m_{\nu_1} m_R}$,
$-\sqrt{m_{\nu_2} m_R}$, $-\sqrt{m_{\nu_3} m_R}$, e.g., \be
O_{\nu}(11) = {\sqrt \frac{\sqrt{m_{\nu_2}}
    \sqrt{m_{\nu_3}}
(D_{\nu}+\sqrt{ m_{\nu_2}} + \sqrt{m_{\nu_3}} )}
{(-\sqrt{m_{\nu_1}}+\sqrt{m_{\nu_2}} + \sqrt{m_{\nu_3}}+ D_{\nu})
(\sqrt{m_{\nu_1}}+\sqrt{m_{\nu_3}}) (\sqrt{m_{\nu_1}} +
\sqrt{m_{\nu_2}} )} } \label{omajih}. \ee The other elements of
diagonalizing transformations in the case of neutrinos as well as
charged leptons can similarly be found.

\section{Inputs used in present calculations\label{inputs}}
Before getting into the details of the analysis, we would like to
mention some of the essentials pertaining to various inputs. The
present work uses results from the latest global three neutrino
oscillation analysis carried out by Fogli {\it et al}.
\cite{inputs}. At 1$ \sigma $ C.L. the allowed ranges of the
various input parameters are
\be
 \Delta {\it m}_{21}^{2} = (7.32-7.80)\times
 10^{-5}~\rm{eV}^{2},~~~~
 \Delta {\it m}_{23}^{2} = (2.33-2.49)\times 10^{-3}~ \rm{eV}^{2},
 \label{deltamass1}\ee
\be
s^2 _{12}  =  (0.29-0.33),~~~
 s^2_{23}  =  (0.37-0.41),~~~
s^2 _{13} = (0.021-0.026), \label{angles1} \ee where $ \Delta {\it
m}_{ij}^{2}$ 's correspond to the solar and atmospheric neutrino
mass square differences and $ s_{ij}$ corresponds to the sine of
the mixing angle $ij$ where $i,j=1,2,3$. At 3$ \sigma $ C.L. the
allowed ranges are given as

\be
 \Delta {\it m}_{21}^{2} = (6.99-8.18)\times
 10^{-5}~\rm{eV}^{2},~~~~
 \Delta {\it m}_{23}^{2} = (2.19-2.62)\times 10^{-3}~ \rm{eV}^{2},
 \label{deltamass2}\ee
\be
s^2 _{12}  =  (0.26-0.36),~~~
 s^2_{23}  =  (0.33-0.64),~~~
s^2 _{13} = (0.017-0.031). \label{angles2} \ee
          For the purpose
of calculations, we have taken the lightest neutrino mass, the
phases $\phi_1$, $\phi_2$ and $D_{l, \nu}$ as free parameters, the
other two masses are constrained by $ \Delta m_{12}^2 =
m_{\nu_2}^2 - m_{\nu_1}^2 $ and $\Delta m_{23}^2 = m_{\nu_3}^2 -
m_{\nu_2}^2 $ in the normal hierarchy case and by $\Delta m_{23}^2
= m_{\nu_2}^2 - m_{\nu_3}^2$ in the inverted hierarchy case. It
may be noted that lightest neutrino mass corresponds to $
m_{\nu_1}$ for the normal hierarchy case and to $m_{\nu_3}$ for
the inverted hierarchy case. For all the three hierarchies of
neutrino masses, the explored range of the lightest neutrino mass
is taken to be $10^{-8}\,\rm{eV}-10^{-1}\,\rm{eV}$, however our
conclusions remain unaffected even if the range is extended
further. The mixing matrix U, given in equation (7) can be
constructed in terms of the elements of the transformations $ O_k
$ and $O_{\nu} $ as well as in terms of the phases  $\phi_1$ and
$\phi_2$. In the absence of CP violation in the leptonic sector,
phases have been given full variation from 0 to $2\pi$ to obtain
the viable ranges of the mixing angles. In carrying this fit we
have constrained $D_{l, \nu}$, the free parameters, such that
diagonalizing transformations $O_l$ and $O_{\nu}$ always remain
real. This implies, for leptons $-(m_{l_2} - m_{l_1})<D_{l}<
(m_{l_3} + m_{l_2})$,  for Majorana neutrinos $-(\sqrt{m_{\nu_2}}
-\sqrt{m_{\nu_1}}) < D_{\nu}
<(\sqrt{m_{\nu_3}} - \sqrt{m_{\nu_2}})$ for normal hierarchy and
$-(\sqrt{m_{\nu_2}} - \sqrt{m_{\nu_1}}) < D_{\nu}
<(\sqrt{m_{\nu_1}} - \sqrt{m_{\nu_3}})$ for inverted hierarchy.

\section{Results and discussion\label{cal}}
\subsection{Inverted hierarchy of neutrino masses \label{inv}}

\begin{figure}[hbt]
  \begin{minipage}{0.45\linewidth}   \centering
\includegraphics[width=2.in,angle=-90]{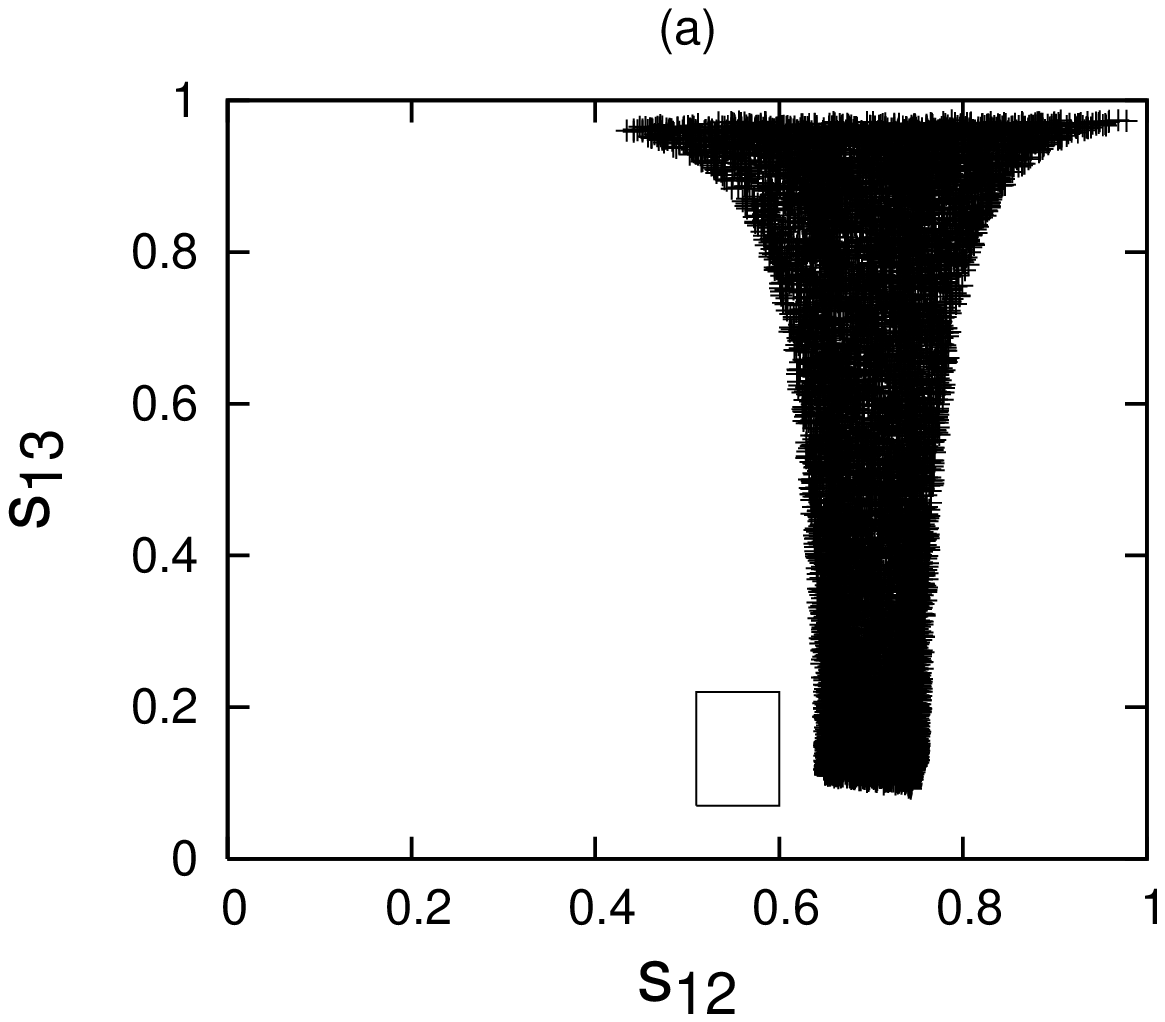}
    \end{minipage} \hspace{0.5cm}
\begin{minipage} {0.45\linewidth} \centering
\includegraphics[width=2.in,angle=-90]{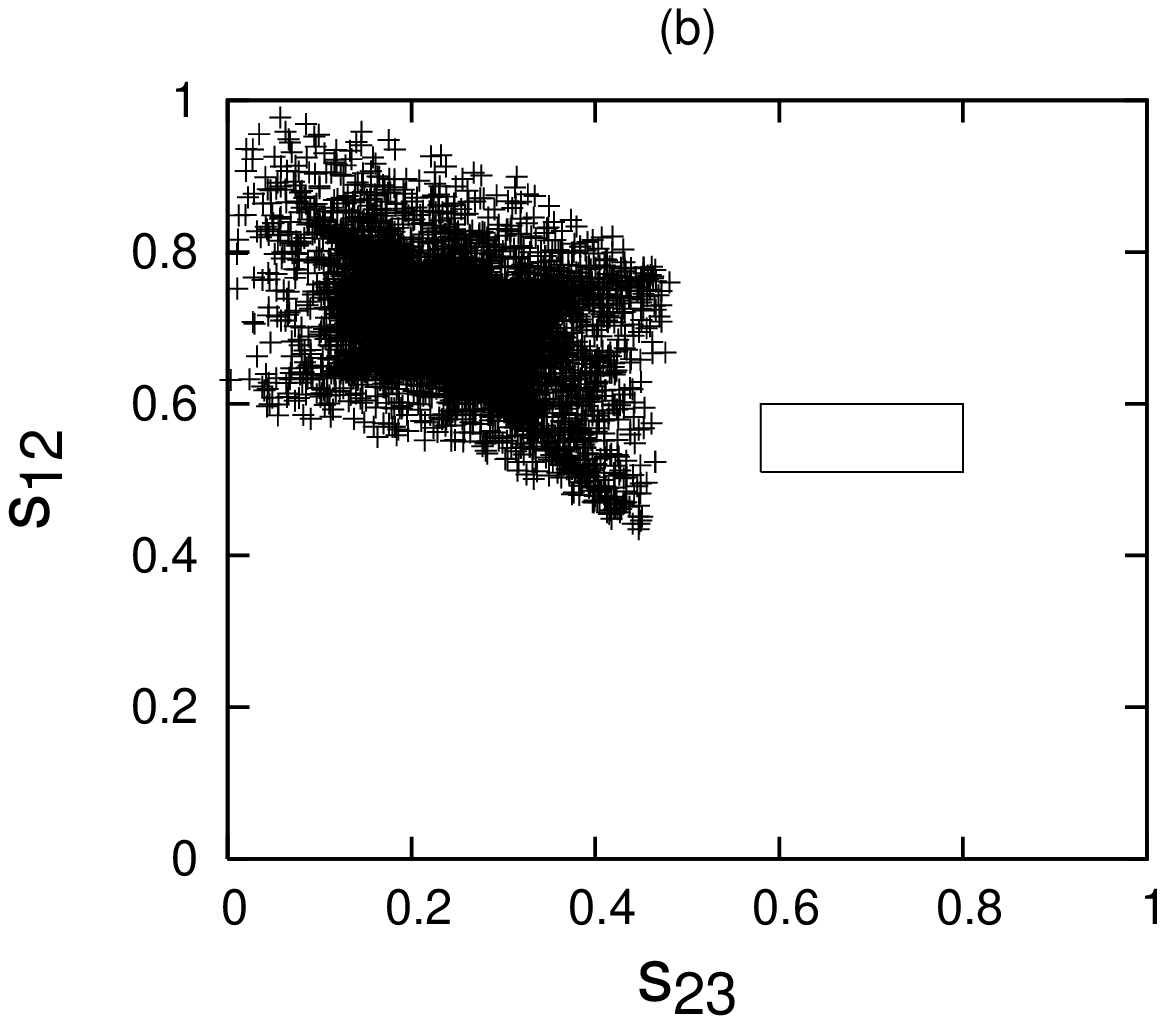}
  \end{minipage}\hspace{0.5cm}
\begin{minipage} {0.45\linewidth} \centering
\includegraphics[width=2.in,angle=-90]{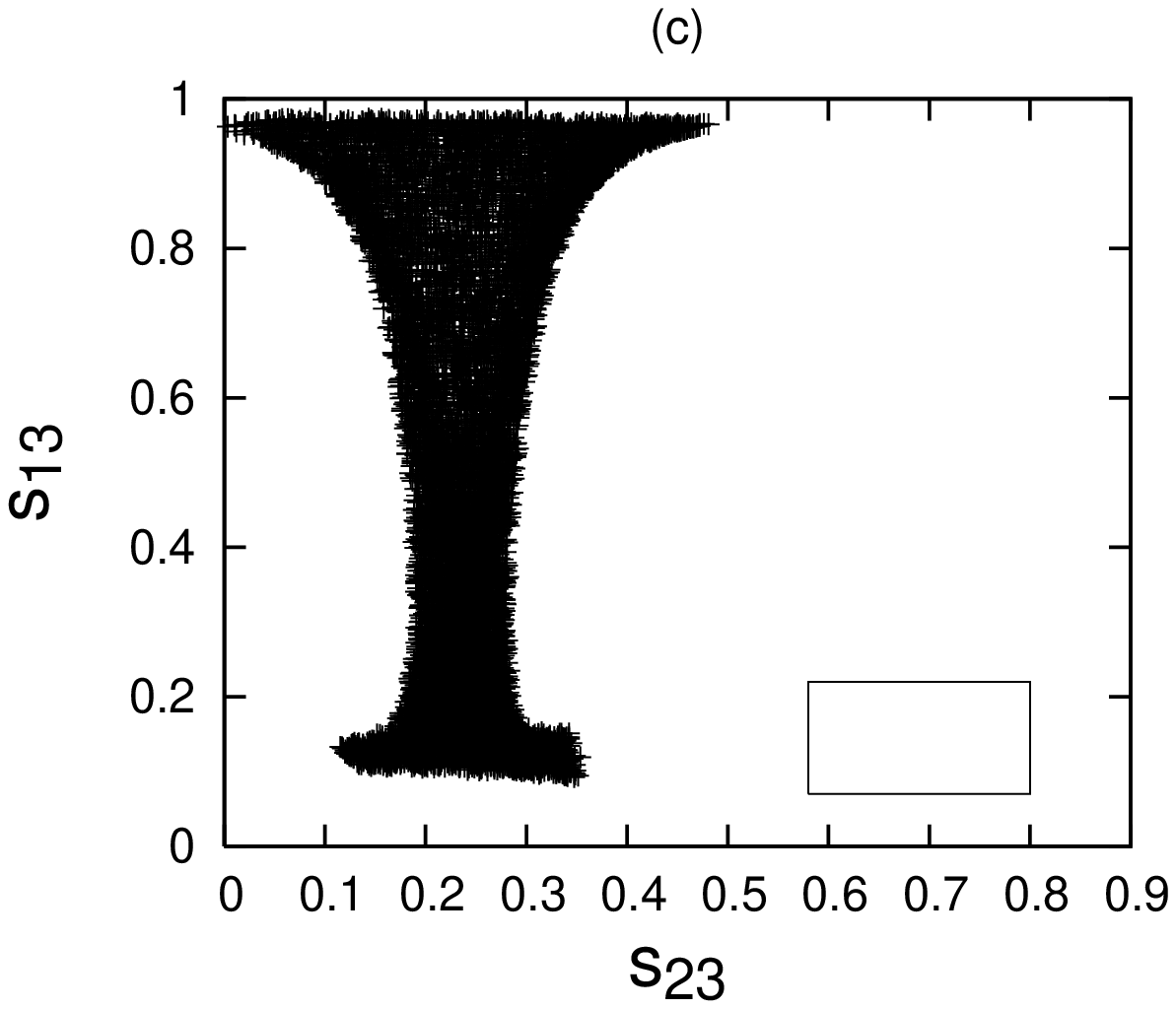}
  \end{minipage}
   \caption{Plots showing the parameter space
corresponding to sines of any of the two mixing angles for texture
6 zero lepton mass matrices. The blank rectangular regions
indicate the experimentally allowed $3\sigma$ C.L. region of the
plotted angles.}
  \label{fig1}
  \end{figure}

To begin with, we discuss the case of inverted hierarchy of
neutrino masses for texture 6 zero as well as texture 5 zero
lepton mass matrices. As a first step, in the case of texture 6
zero mass matrices, we have made an attempt to explicitly show the
ruling out of inverted hierarchy of neutrino masses, thereby
reinforcing the conclusions of  FSTY \cite{fukugita}. To this end
in Figures (\ref{fig1}a), (\ref{fig1}b) and (\ref{fig1}c), we have
plotted the parameter space corresponding to sines of any of the
two mixing angles by constraining the third angle by its
experimental bound given in equation (\ref{angles2}) and giving
full allowed variation to other parameters. Also included in the
figures are blank rectangular regions indicating the
experimentally allowed $3\sigma$ C.L. region of the plotted
angles. Interestingly, a general look at these figures reveals
that inverted hierarchy is clearly ruled out. In particular, from
Figure (\ref{fig1}a) showing the plot of $s_{12}$ versus $s_{13}$,
one can immediately conclude that the plotted parameter space of
these two angles has no overlap with their experimentally allowed
$3\sigma$ C.L. region. This clearly indicates that at 3$\sigma$
C.L. inverted hierarchy of neutrino masses is not viable. These
conclusions are further reinforced from Figures (\ref{fig1}b) and
(\ref{fig1}c) wherein we have plotted $s_{23}$ versus $s_{12}$ and
$s_{23}$ versus $s_{13}$ respectively. Both the figures also
indicate that the plotted parameter space does not include
simultaneously the experimental bounds of the sines of the plotted
angles. Therefore, one can conclude that inverted hierarchy is
ruled out for texture 6 zero neutrino mass matrices.

\begin{figure}
\bc
 \vspace{0.12in}\psfig{file=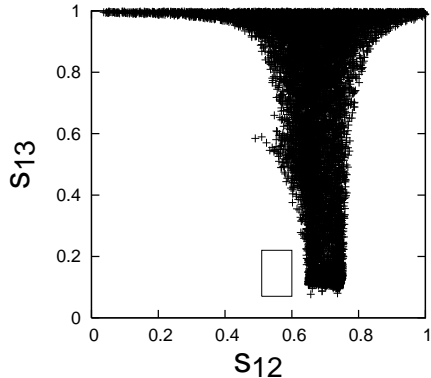, width=2.in,
angle=270}\vspace{0.05in}
 \caption{Plot showing the parameter space corresponding to $ s_{12} $ versus $ s_{13} $
for the $D_l$=0 case of texture 5 zero lepton mass matrices. The
blank rectangular region indicates the experimentally allowed
$3\sigma$ C.L.  region of the plotted angles.} \label{nhih5zdn}
 \ec
\end{figure}

Coming to the two cases of Fritzsch-like texture 5 zero mass
matrices, we first discuss the case when $D_l=0$ and $D_{\nu} \neq
0$. Primarily to facilitate comparison with texture 6 zero case,
in Figure (\ref{nhih5zdn}) we have plotted
 $s _{12} $ against $ s_{13} $ for
 inverted hierarchy for a particular value of $D_{\nu}= \sqrt{m_{\nu_3}}$.
The figure  clearly reveals that inverted hierarchy is ruled out
as again the plotted parameter space has no overlap with the
experimental $3\sigma$ C.L. ranges of  $s _{12} $ and $ s_{13}$.
For the second case of texture 5 zero mass matrices pertaining to
$D_{\nu}=0$ and $D_l \neq 0$ case, one can again plot figures
similar to the ones shown in Figures (\ref{fig1}) and
(\ref{nhih5zdn}). Interestingly, these graphs are essentially
similar as shown in Figure (\ref{fig1}), therefore these are not
presented here. This can be understood by the fact that there is a
very strong hierarchy in the case of charged leptons which reduces
the texture 5 zero $D_{\nu}=0$ case essentially to texture 6 zero
case only. By similar arguments, this case can also be therefore
ruled out for inverted hierarchy of neutrino masses.

\subsection{Degenerate scenario of neutrino masses \label{deg}}

\begin{figure}[hbt]
  \begin{minipage}{0.45\linewidth}   \centering
\includegraphics[width=2.in,angle=-90]{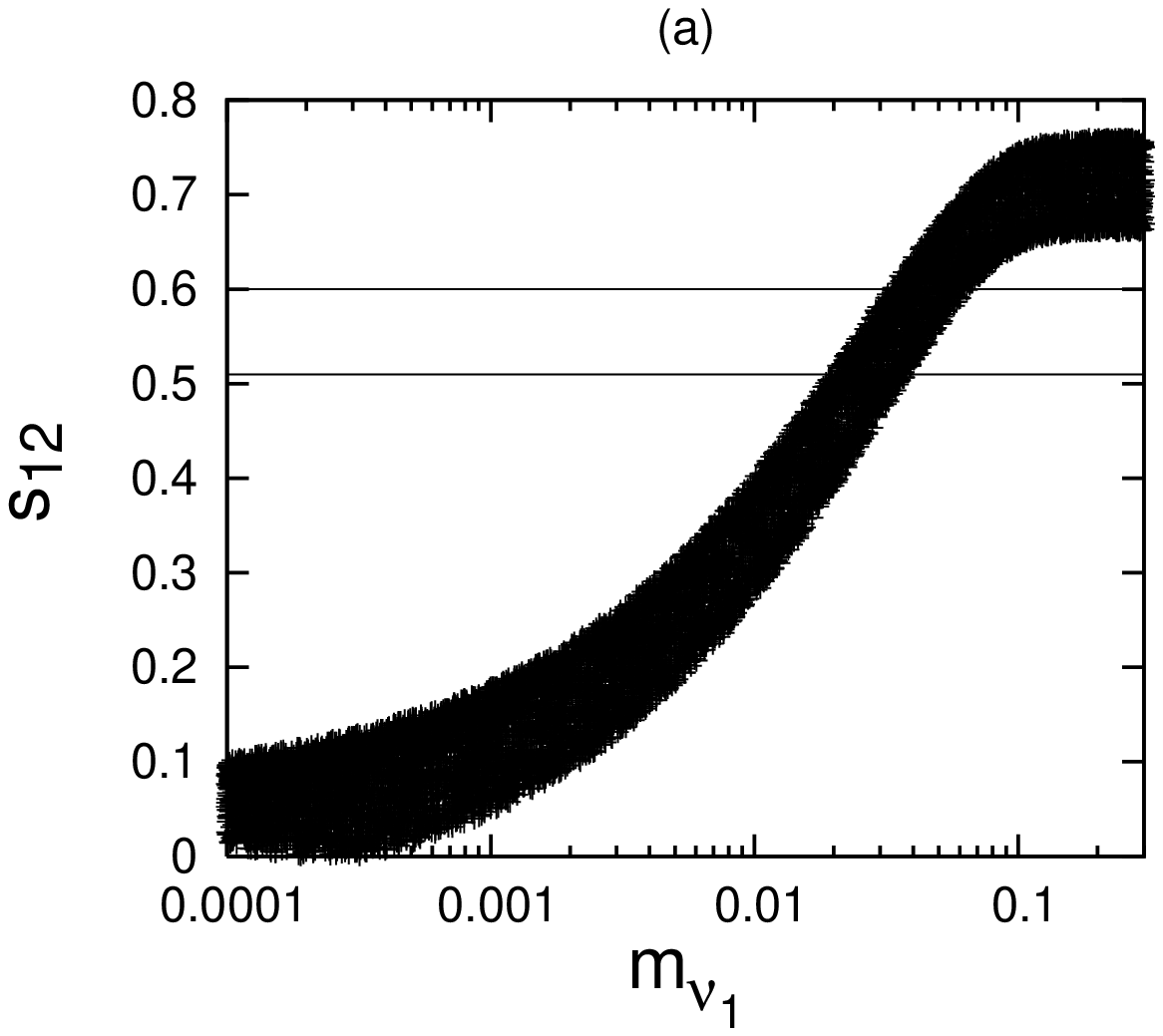}
    \end{minipage} \hspace{0.5cm}
     \begin{minipage}{0.45\linewidth}   \centering
\includegraphics[width=2.in,angle=-90]{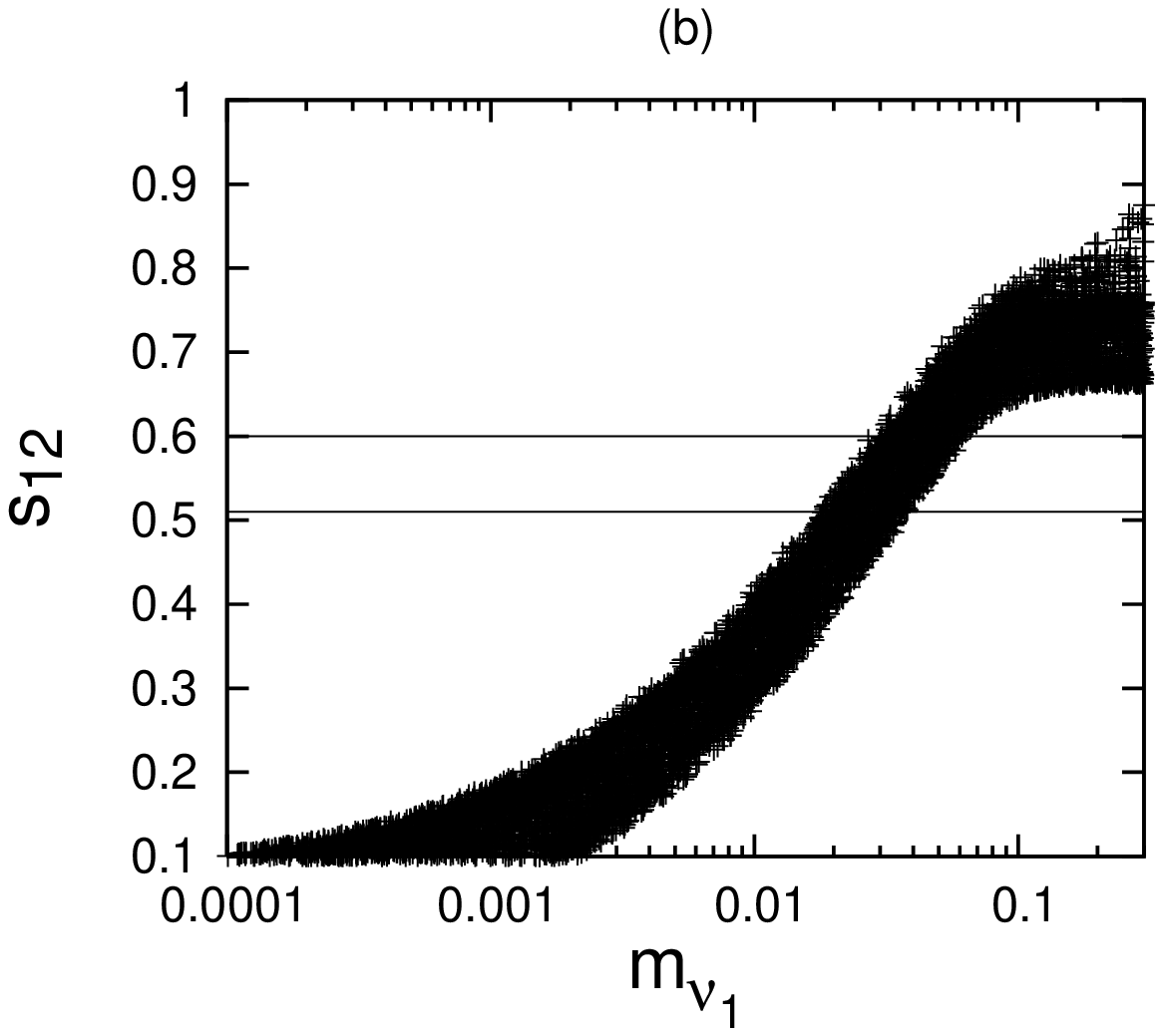}
    \end{minipage} \hspace{0.5cm}
 \caption{Plot showing the variation of
$s_{12}$ with lightest neutrino mass $m_{\nu_1}$ for (a) texture 6
zero lepton mass matrices and (b) $D_l$=0 case of texture 5 zero
lepton mass matrices. The parallel lines indicate the $3\sigma$
C.L. limits of $s_{12}$.} \label{degm1}
\end{figure}

The cases of neutrino masses being degenerate, characterized by
either $m_{\nu_1} \lesssim m_{\nu_2} \sim m_{\nu_3} \lesssim
0.1~\rm{eV}$ or $m_{\nu_3} \sim m_{\nu_1} \lesssim m_{\nu_2}
\lesssim 0.1~\rm{eV}$ corresponding to normal and inverted
hierarchy respectively, can again be shown to be ruled out. For
texture 6 zero and 5 zero mass matrices considering degenerate
scenario corresponding to inverted hierarchy of neutrino masses,
Figures (\ref{fig1}) and (\ref{nhih5zdn}) can again be used to
rule it out at 3$\sigma$ C.L.. While plotting these figures the
range of the lightest neutrino mass is taken to be
$10^{-8}\,\rm{eV}-10^{-1}\,\rm{eV}$, which includes the neutrino
masses corresponding to degenerate scenario, therefore by
discussion similar to the one given for ruling out inverted
hierarchy and degenerate scenario of neutrino masses is ruled out
as well.

Coming to the degenerate scenario corresponding to normal
hierarchy of neutrino masses, one can  show that this is ruled out
as well. To this end, for texture 6 zero mass matrices in Figure
(\ref{degm1}a) we have plotted the mixing angle $s_{12}$ against
the lightest neutrino mass $m_{\nu_1}$. From the figure one can
find that the values of $s_{12}$ corresponding to $m_{\nu_1}
\lesssim 0.1~\rm{eV}$ lie outside the experimentally allowed
range, thereby ruling out degenerate scenario. For the texture 5
zero mass matrices, pertaining to $D_l=0$ case, the degenerate
scenario corresponding to normal hierarchy  of neutrino masses can
again be shown to be ruled out. Similar to the case of texture 6
zero mass matrices Figure (\ref{degm1}b) again shows that the
values of $s_{12}$ corresponding to $m_{\nu_1} \lesssim
0.1~\rm{eV}$ lie outside the experimentally allowed range. For the
$ D_{\nu}$=0 case of texture 5 zero mass matrices
 again one gets essentially the same figures
as for the case of texture 6 zero mass matrices therefore
degenerate scenario for this case can be ruled out by similar
arguments.

\subsection{Normal hierarchy of neutrino masses \label{nor}}
\subsubsection{ Texture 6 zero lepton mass matrices}
To begin with, we have made an attempt to reproduce the results of
FSTY \cite{fukugita} for texture 6 zero lepton mass matrices.
Using the latest inputs we are largely able to reproduce their
results. For the sake of comparison, we have presented our results
pertaining to effective Majorana mass measured in neutrinoless
double beta decay $(\beta\beta)_{0 \, \nu}$, given as
\be
\langle m_{ee} \rangle = m_{\nu_1} U_{e1}^2 + m_{\nu_2} U_{e2}^2
+m_{\nu_3} U_{e3}^2. \label{m6} \ee\begin{figure} \bc
\vspace{0.12in}\psfig{file=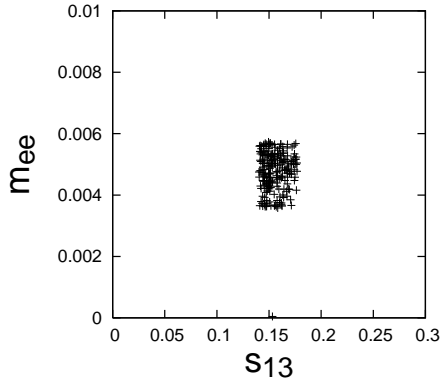, width=2.in,
angle=270}\vspace{0.08in} \caption{Plot showing variation of
effective Majorana mass measured in neutrinoless double beta decay
$(\beta\beta)_{0 \, \nu}$ (in eV) with the mixing angle $ s_{13} $
for texture 6 zero lepton mass matrices.} \label{mee6} \ec
\end{figure}
In particular, in Figure (\ref{mee6}) we have plotted the
variation of $\langle m_{ee} \rangle$ with respect to the mixing
angle $ s_{13} $. From the figure it is clear that there is a good
overlap between the present calculations and the results of FSTY.
\begin{figure}[hbt]
  \begin{minipage}{0.45\linewidth}   \centering
\includegraphics[width=2.in,angle=-90]{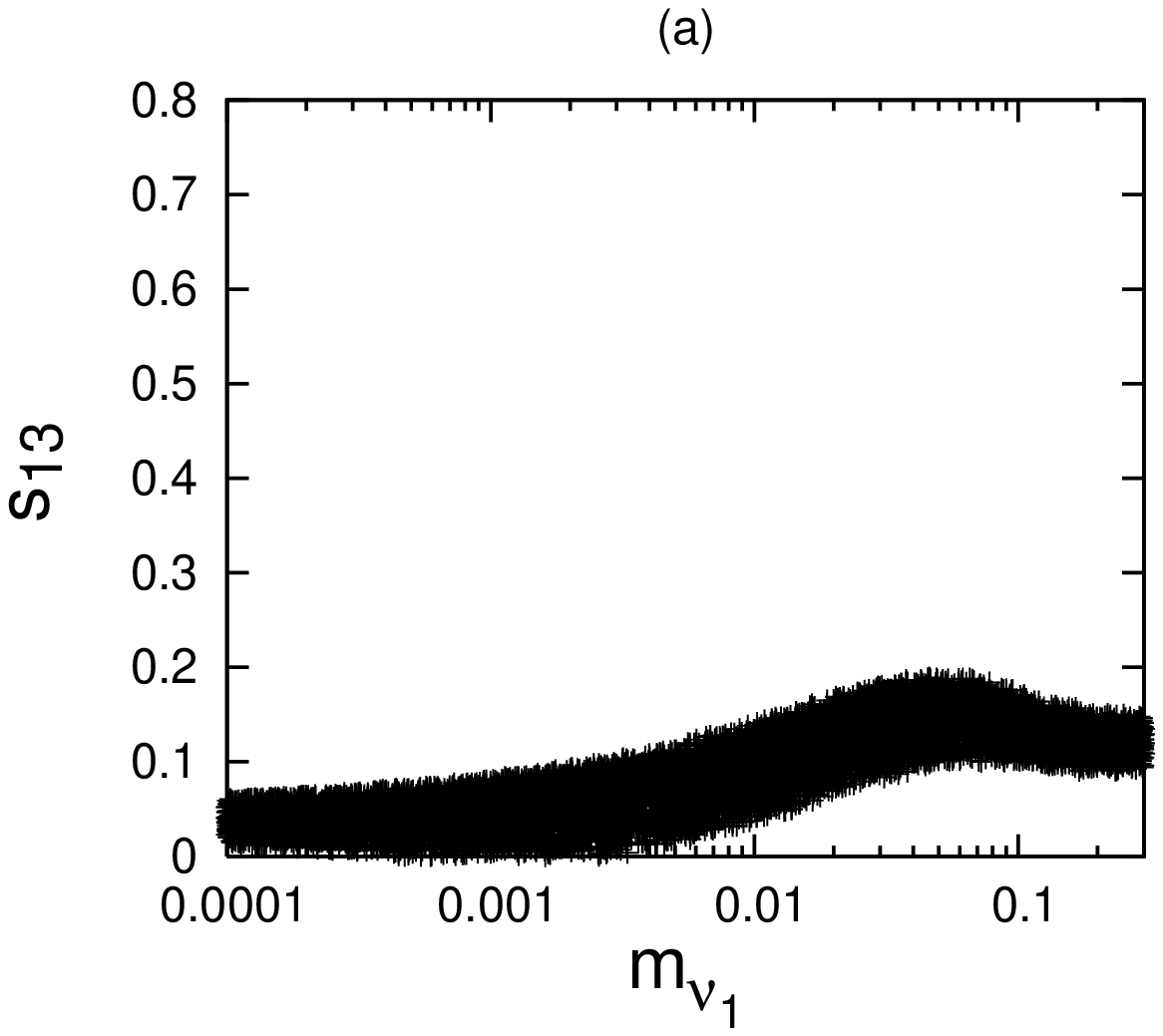}
    \end{minipage} \hspace{0.5cm}
\begin{minipage} {0.45\linewidth} \centering
\includegraphics[width=2.in,angle=-90]{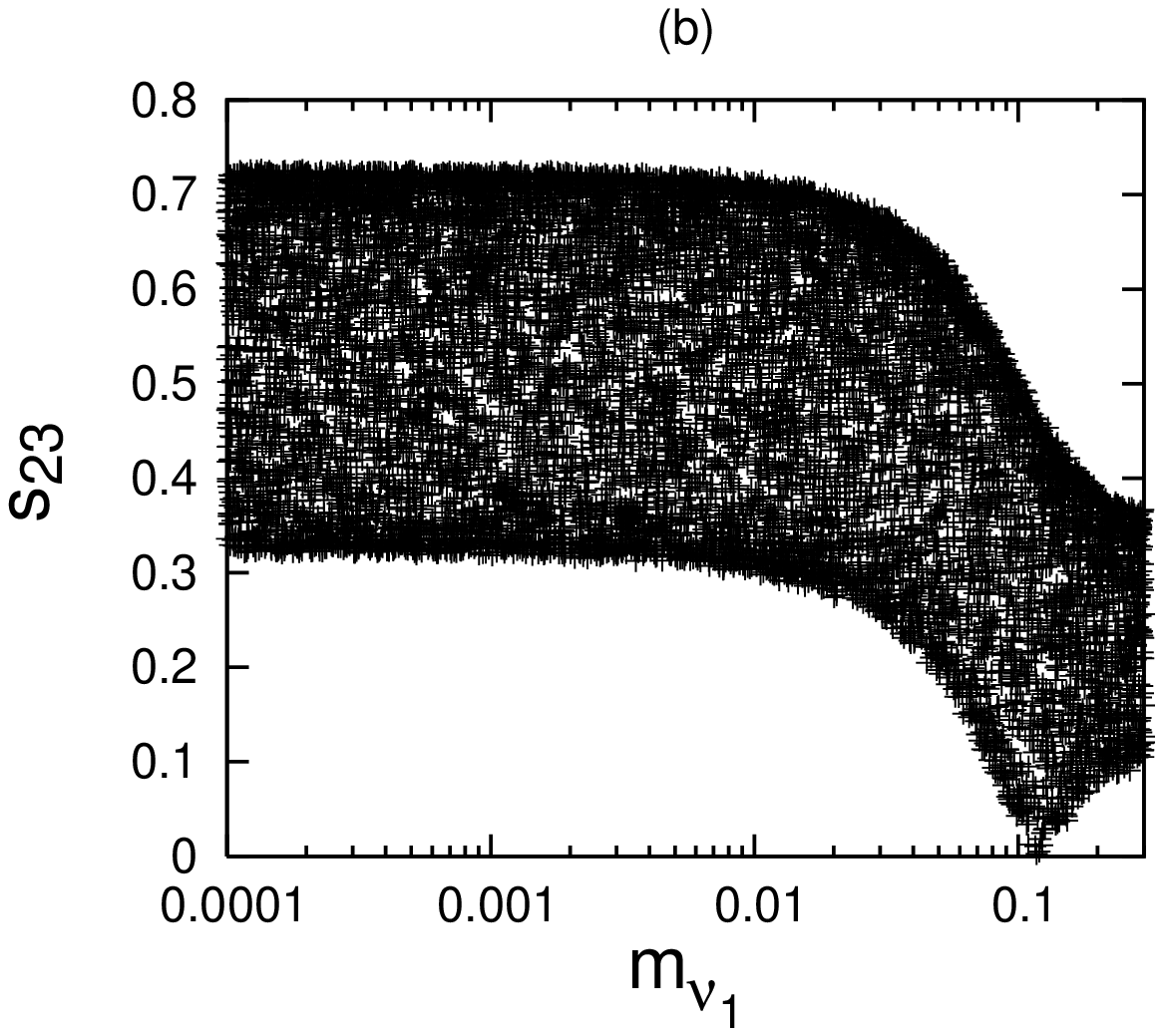}
\end{minipage}\hspace{0.5cm}
\caption{Plots showing variation of the mixing angles $s_{13}$ and
$s_{23}$ with the lightest neutrino mass $ m_{\nu_1} $ for texture
6 zero lepton mass matrices.} \label{nhih6z}
\end{figure}

As an extension of the analysis of FSTY for the texture 6 zero
lepton mass matrices for normal hierarchy of neutrino masses, in
Figures (\ref{degm1}a) and \ref{nhih6z}(a, b) we have respectively
plotted the lightest neutrino mass $ m_{\nu_1}$ against the mixing
angles $ s_{12}, s_{13}$ and $ s_{23} $. While plotting any of
these graphs, the other input parameters have been given
variations at $3\sigma $ C.L.. It may be noted that the plot
between $ s_{23}$ and $ m_{\nu_1}$, Figure (\ref{nhih6z}b)
expectedly does not provide any constrains on the range of $
m_{\nu_1}$ considered here. Similarly, considering the $3\sigma $
C.L. range of $ s_{13} $ we do not get any reasonable constrains
on $ m_{\nu_1}$, however, interestingly because of the almost
precise value of $ s_{12} $ even at $3\sigma $ C.L. we get a
narrow range of the lightest neutrino mass $ m_{\nu_1}$=0.03-0.07
eV. From Figures (\ref{degm1}a) and (\ref{nhih6z}a) one finds that
for 1$\sigma$ C.L. experimental ranges of $s_{12}$ and  $ s_{13}$
we get the range of  $ m_{\nu_1}$ to be 0.03-0.06 eV and 0.007-0.3
eV respectively. It may be noted that any measurement regarding
the lightest neutrino mass $ m_{\nu_1}$ will have immediate
implications for texture 6 zero mass matrices considered here.

\subsubsection{Texture 5 zero mass matrices}
In the case of texture 5 zero lepton mass matrices, we have
carried out calculations for both the cases  when $D_l=0$ with
$D_{\nu} \neq 0$ and $D_{\nu}=0$ with $D_l \neq 0$. Since the
texture 6 zero mass matrices are able to accommodate the lepton
mixing data quite well, it is therefore expected that both the
cases of texture 5 zero mass matrices would also be viable.
Similar to the case of texture 6 zero mass matrices, we would like
to study the implications of the mixing angles on the lightest
neutrino mass $ m_{\nu_1}$, in particular that of $ \theta_{13} $.

\begin{figure}[hbt]
 \begin{minipage} {0.45\linewidth} \centering
\includegraphics[width=2.in,angle=-90]{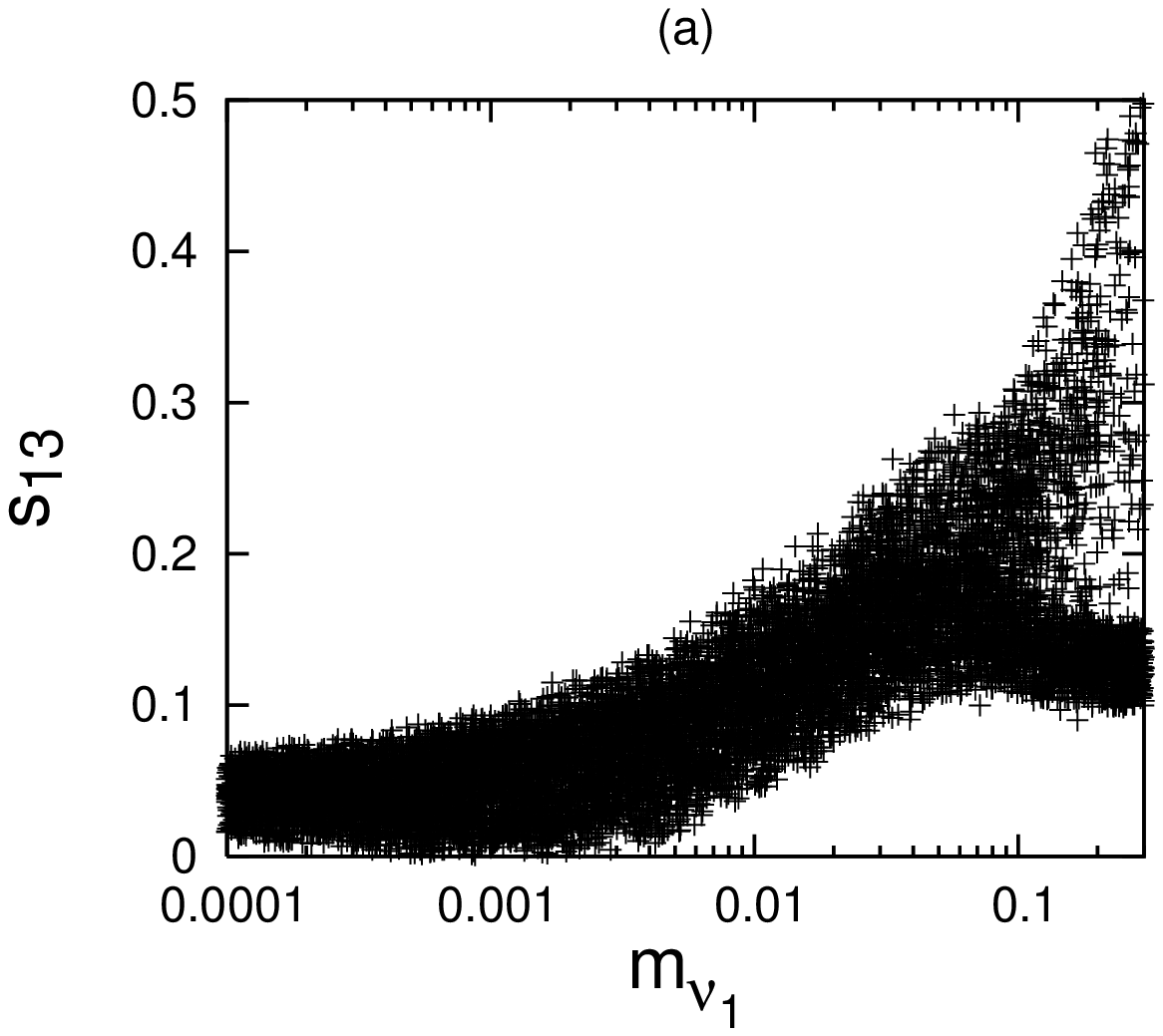}
  \end{minipage}\hspace{0.5cm}
\begin{minipage} {0.45\linewidth} \centering
\includegraphics[width=2.in,angle=-90]{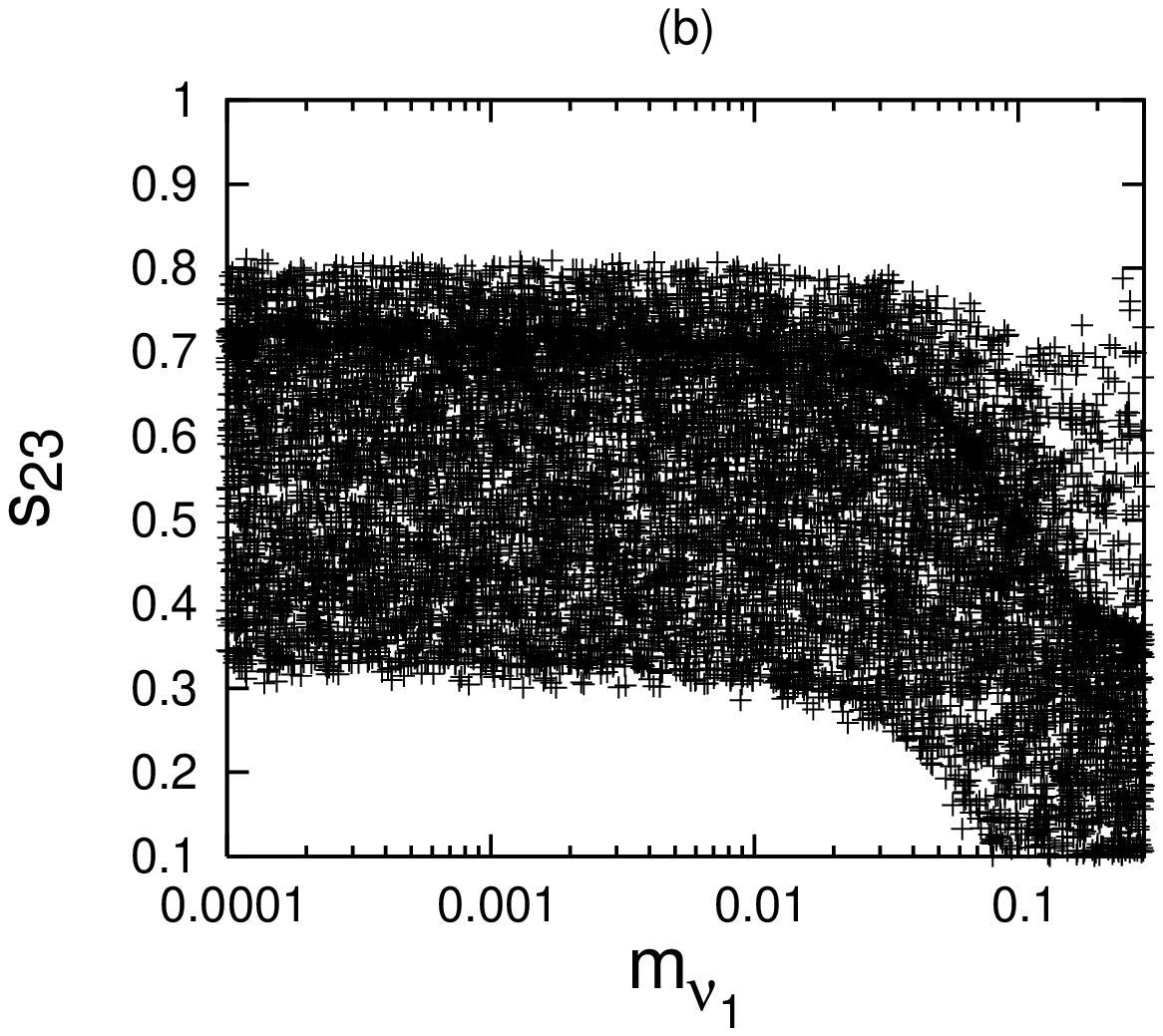}
  \end{minipage}
 \caption{Plots showing variation of
mixing angles  $s_{13}$ and $s_{23}$ with the lightest neutrino
mass $ m_{\nu_1} $  for texture 5 zero lepton mass matrices when
$D_l=0$ and $D_{\nu} \neq 0$.} \label{nhih5z}
\end{figure}

We first discuss the case when $D_l=0$ and $D_{\nu} \neq 0$.  For
normal hierarchy of neutrino masses, the additional parameter $
D_{\nu}$ can be constrained such that the diagonalizing
transformations $ O_l$ and $O_{\nu} $ always remain real.
Primarily to facilitate comparison with texture 6 zero case, in
Figures (\ref{degm1}b) and (\ref{nhih5z}) we have plotted the
mixing angles against the lightest neutrino mass for normal
hierarchy of neutrino masses.  Interestingly, we find that this case shows a big change in the
behaviour of the mixing angles $ s_{13}$ and $ s_{23} $ versus the
lightest neutrino mass as compared to the texture 6 zero graphs
shown in Figure (\ref{nhih6z}). It may be noted that similar to
the case of texture 6 zero mass matrices, $ s_{13}$ and $ s_{23} $
are not able to constrain the lightest neutrino mass $ m_{\nu_1}$
at $3\sigma $ C.L.. However at $1\sigma $ C.L., $ s_{13}$
constrains the range of $ m_{\nu_1}$ to be 0.007-0.03 eV. Similar
to the case of texture 6 zero mass matrices, the present refined
value of $ s_{12}$, even at $3\sigma $ C.L., provides the range of
the lightest neutrino mass $ m_{\nu_1}$ as 0.03-0.08 eV, which is
 somewhat expanded in
the present case as compared  to texture 6 zero mass matrices.
This can be clearly understood by noting the fact that $D_l=0$ and
$D_{\nu} \neq 0 $ case introduces additional parameters. It may be
added that on comparison with our earlier analysis \cite{FLMM}
which was carried out when only an upper bound on $s_{13}$ was
known, we find that the present experimental range of $s_{13}$ has
sharpened the range of the lightest neutrino mass $ m_{\nu_1} $.

 \begin{figure}[hbt]
\bc \vspace{0.12in}\psfig{file=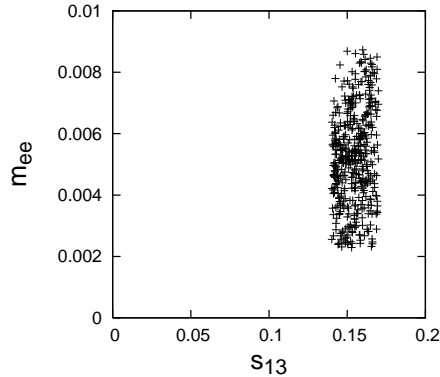, width=2.in,
angle=270}\vspace{0.08in}
 \caption{Plot showing variation of
 effective Majorana mass measured in
neutrinoless double beta decay $(\beta\beta)_{0 \, \nu}$ (in eV)
 with the mixing angle $ s_{13} $ for texture 5 zero lepton mass matrices. } \label{mee5}\ec
\end{figure}

As a next step, similar to the case of texture 6 zero mass
matrices, in Figure (\ref{mee5}) we have shown the variation of
the mixing angle $ s_{13} $ against the effective Majorana mass
$\langle m_{ee} \rangle$. One finds that the $1\sigma $ C.L. range
of the mixing angle $ \theta_{13} $ constrains the value of
$\langle m_{ee} \rangle$  to be 2.3-8.7 meV, this being broadly in
agreement with \cite{bilenky}. Due to the additional parameter $
D_{\nu} $ the range obtained here also shows an expansion in
comparison to the one obtained in the case of texture 6 zero mass
matrices. It may be of interest to mention that the Jarlskog's
rephasing invariant parameter in the leptonic sector $J_l$
\cite{jarlskog} spans the range
 -4.23 $ \times 10^{-2}-4.36 \times
 10^{-2}$.

\begin{figure}[hbt]
  \begin{minipage}{0.45\linewidth}   \centering
\includegraphics[width=2.in,angle=-90]{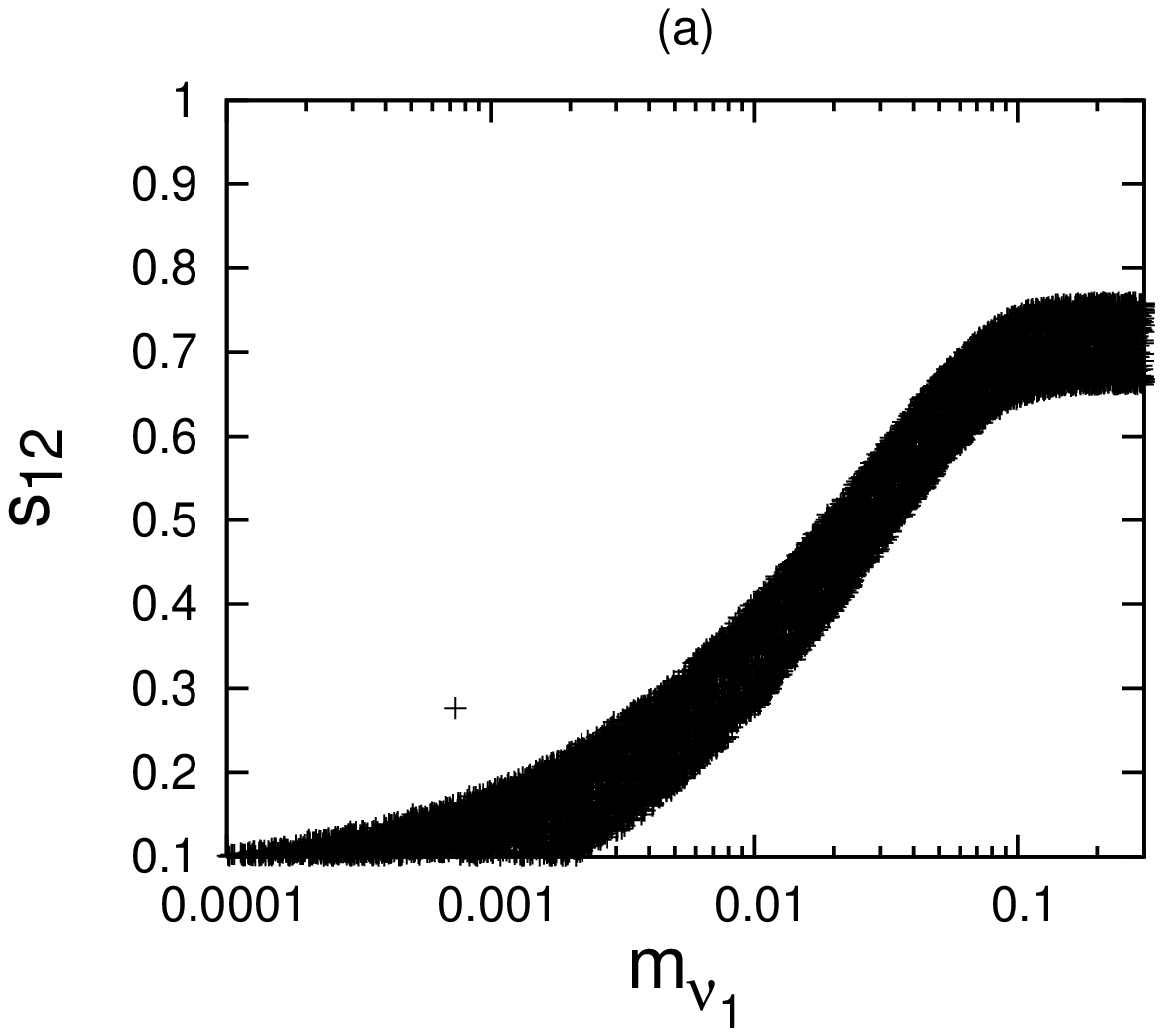}
    \end{minipage} \hspace{0.5cm}
\begin{minipage} {0.45\linewidth} \centering
\includegraphics[width=2.in,angle=-90]{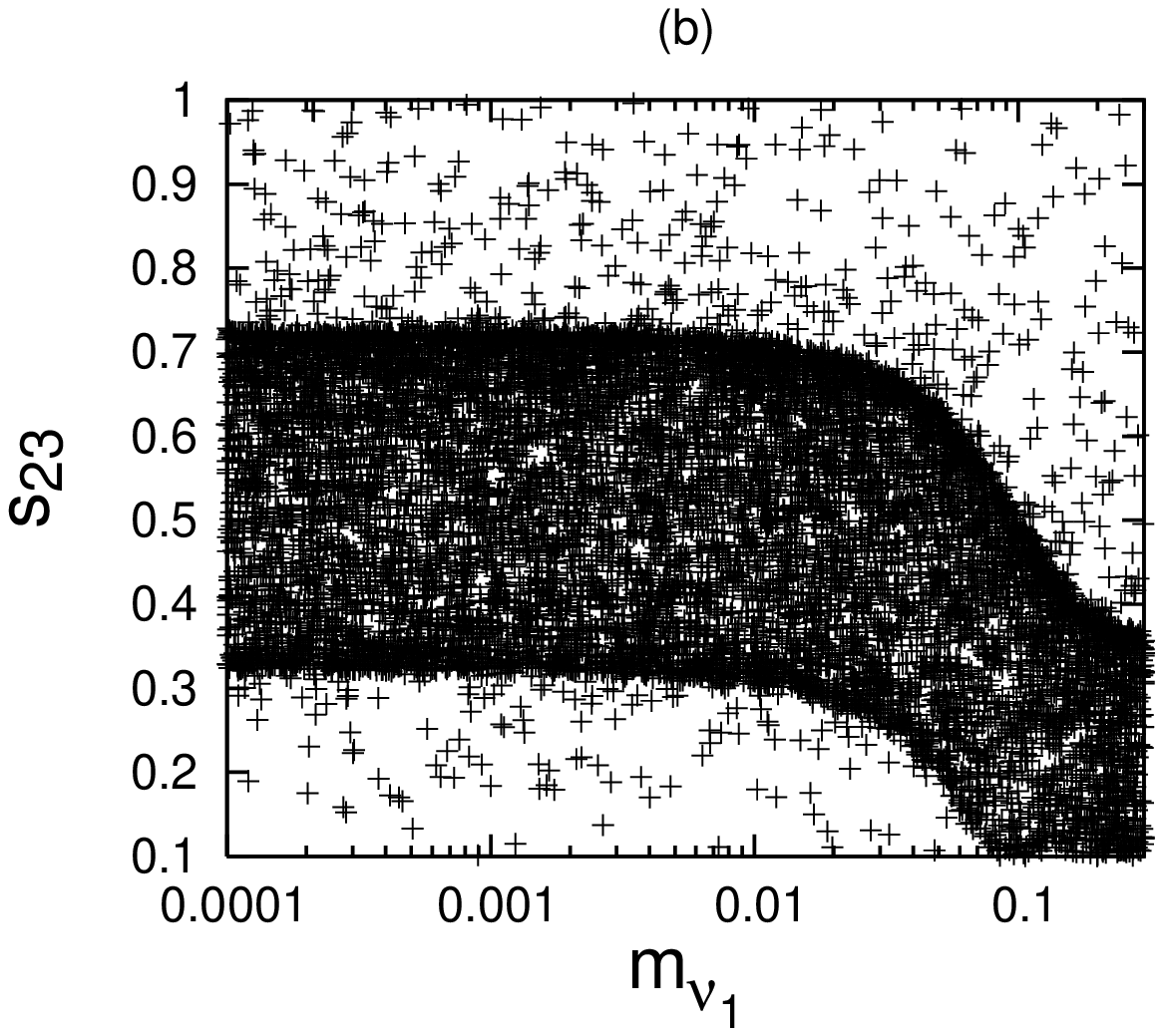}
  \end{minipage}\hspace{0.5cm}
\begin{minipage} {0.45\linewidth} \centering
\includegraphics[width=2.in,angle=-90]{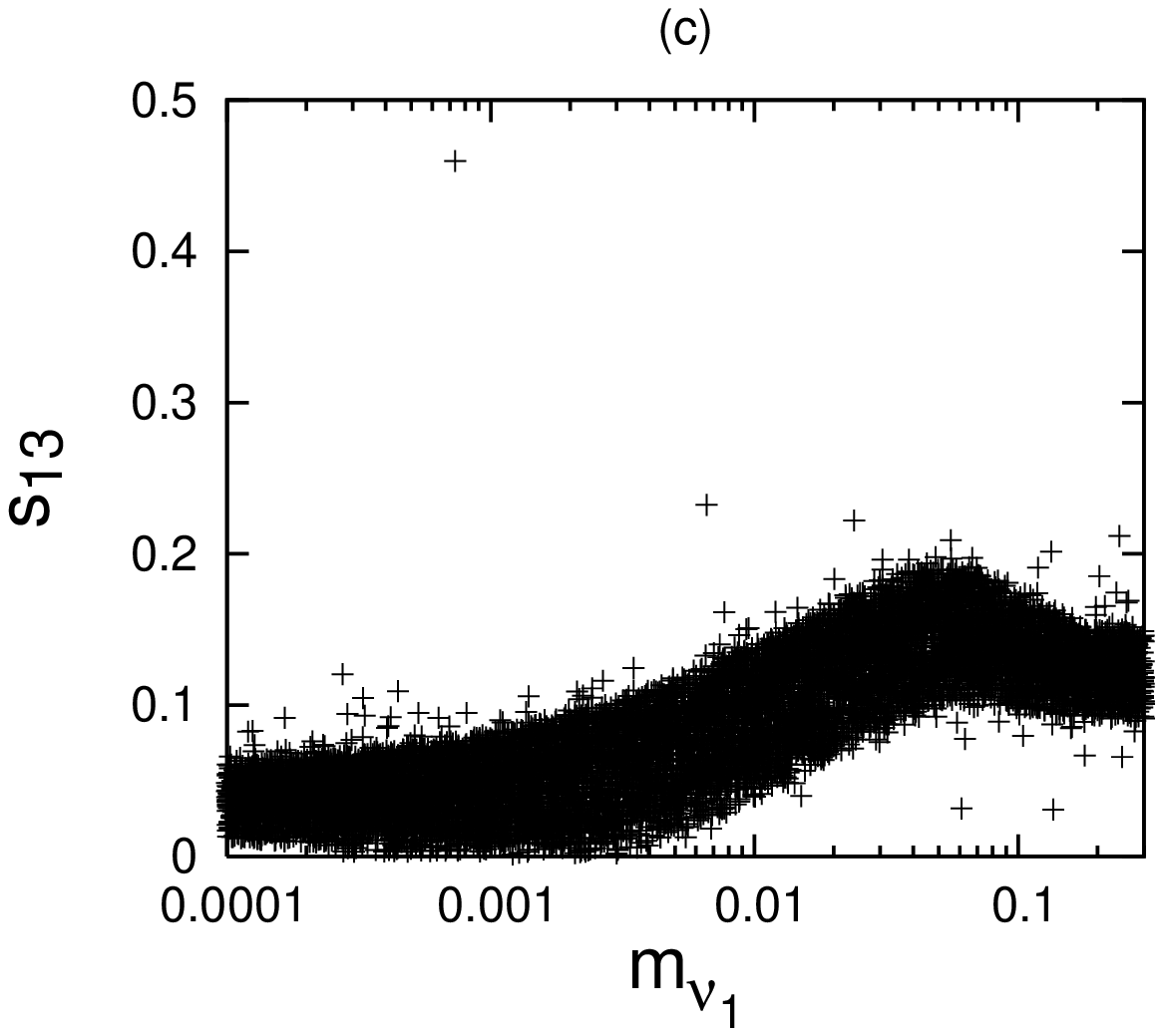}
  \end{minipage}
 \caption{Plots showing variation of
mixing angles $s_{12}$, $s_{13}$ and $s_{23}$ with the lightest
neutrino mass $ m_{\nu_1} $  for texture 5 zero lepton mass
matrices  when $D_l\neq 0$ and $D_{\nu} = 0$. } \label{nhih55z}
\end{figure}

Coming to the second case of  texture 5 zero mass matrices where
$D_{\nu}=0$ and $D_l \neq 0$, the plots of the  mixing angles
against the lightest neutrino mass are shown in Figure
(\ref{nhih55z}). Interestingly these graphs are very similar to
those in Figure (\ref{nhih6z}) corresponding to the texture 6 zero
case. This can again be understood by noting the fact that there
is very strong hierarchy in the case of charged leptons which
reduces the texture 5 zero $D_{\nu}=0$ case essentially to the
texture 6 zero case. Also we would like to mention  that out of
the two free parameters $D_l$ and $D_{\nu}$, the parameter
$D_{\nu}$ plays a more important role than $D_l$. This becomes
more clear if we compare the graphs of the texture 6 zero matrices
with those of the two cases of the texture 5 zero matrices. The
texture 6 zero plots and those of the texture 5 zero $D_{\nu}=0$
case are essentially similar whereas those of the texture 5 zero
$D_l=0$ case, wherein $D_{\nu}$ is non zero, are different from
the graphs of the texture 6 zero matrices.

In this case also we present the viable range of Jarlskog's
rephasing invariant parameter in the leptonic sector $J_l$ which
is predicted to be -4.11 $ \times 10^{-2}-4.26 \times 10 ^2 $. The
$\langle m_{ee} \rangle$ range comes out to be 2.4 - 6.02 meV
which includes the range obtained in texture 6 zero case given as
3.7 - 5.6 meV. In the absence of any definite information about
$J_l$ and $\langle m_{ee} \rangle$ we find that the ranges
corresponding to different cases are in broad agreement with the
similar calculations done by several other authors \cite{bilenky}.

\section{Comments, summary and conclusions\label{summ}}
Before summarizing the present work, a few comments are in order.
It may be mentioned that in the case when charged leptons are in
the flavor basis, the mixing matrix becomes much more simplified
and one can easily deduce the consequences for different
hierarchies of neutrino masses for the texture 6 zero as well as
texture 5 zero mass matrices. Further, it may be added that the
ranges of $ D_{\nu} $ and $ D_l $ taken here, suggest that the
present mass matrices can be considered as `natural' as advocated
by Peccei and Wang \cite{natural}. This analysis also brings forth
an important point that even though the neutrino mixing pattern is
very different from the quark mixing pattern, yet both can be
described by `natural' texture specific mass matrices.

To summarize, as a first step we have extended the very recent
analysis of FSTY \cite{fukugita} regarding compatibility of
texture 6 zero hermitian lepton mass matrices with the leptonic
mixing data. Apart from reproducing their results we have been
able to clearly rule out inverted hierarchy and degenerate
scenario of neutrino masses in texture 6 zero mass matrices.
Further, keeping in mind quark-lepton universality \cite{smirnov},
in the present work we have carried out the analysis for the
texture 5 zero  lepton mass matrices also as these are not ruled
out in the case of quarks, unlike the case of texture 6 zero
Fritzsch-like and non Fritzsch-like quark mass matrices.
Interestingly, again inverted hierarchy and degenerate scenario of
neutrino masses for texture 5 zero mass matrices also seem to be
ruled out.

The present analysis indicates that the present $3\sigma $ C.L.
range of $\theta_{13}$ does not put a reasonable constraint on the
value of lightest neutrino mass $m_{\nu_1} $, therefore,
refinements of its value will have important implications for $
m_{\nu_1}$. Regarding, the effective Majorana mass $\langle m_{ee}
\rangle$, one finds that for texture 5 zero lepton mass matrices
when $D_l=0$ and $D_{\nu} \neq 0$ the $1\sigma $ C.L. range of the
mixing angle $ \theta_{13} $ constrains the value of $\langle
m_{ee} \rangle$  to be 2.3-8.7 meV. This range looks to be
somewhat expanded in comparison to the one obtained in the case of
texture 6 zero mass matrices due to the additional parameter $
D_{\nu} $. Therefore, it seems that measurements of  $ m_{\nu_1}$
and $\langle m_{ee} \rangle$ would have important implications on
texture specific mass matrices considered here.

\vskip 0.2cm
{\bf Acknowledgements} \\ P.F. would like to thank
University Grants Commission (Ref. No: F.
4-3/2006(BSR)/5-89/2007(BSR)) for financial support. G.A. would
like to acknowledge DST, Government of India (Grant No:
SR/FTP/PS-017/2012) for financial support. P.F., S.S., G.A.
acknowledge the Chairman, Department of Physics, P.U., for
providing facilities to work. R.V. would like to thank the
Director, RIEIT for providing adequate facilities.

\end{document}